\documentclass[twocolumn,prl,showpacs]{revtex4}
\usepackage{graphicx}
\usepackage{epsfig}
\usepackage{bm}

\begin{document}

\title{Phase Diagrams of Quasispecies Theory with Recombination
and Horizontal Gene Transfer}

\author{J.-M. Park$^{1,2}$
and M.\ W.\ Deem$^1$
}

\affiliation{
\hbox{}$^1$Department of Physics \& Astronomy,
Rice University, Houston,Texas 77005--1892, USA\\
\hbox{}$^2$Department of Physics,
The Catholic University of Korea,
Bucheon, 420--743, Korea\\
}

\begin{abstract}
We consider how transfer of genetic information between
individuals influences the phase diagram and mean fitness
of both the Eigen and
the parallel, or Crow-Kimura,  models of evolution.  In the absence of
genetic transfer, these physical models of evolution consider
the replication and point mutation of the genomes of independent
individuals in a large population.  A phase transition occurs, such that below
a critical mutation rate an identifiable quasispecies forms.
We show how transfer of genetic information changes the 
phase diagram  and mean fitness
and introduces metastability in quasispecies theory,
via an analytic field theoretic mapping.
\end{abstract}


\maketitle

We consider how quasispecies evolution changes in the
presence of transfer of genetic
information between individuals in a population.  
That is, we quantify by quasispecies theory
the mutational load, if any, introduced by 
a model of recombination and gene transfer.
Exchange of genetic information between individuals is
believed to be pervasive in nature and crucial to evolutionary
dynamics (for reviews, see \cite{len12,Shapiro3,Otto}).
Experiments and theory have emphasized
that recombination and gene transfer in various forms
increase the rate of laboratory directed
protein evolution \cite{len7,len8,Deem1999} (for reviews see
\cite{len11,Lutz}).
Other experiments have amplified this point and have also
suggested that, while significant
in practice, the advantage of
recombination may simply be to speed up the evolutionary process
that would naturally occur by mutation alone in the limit
of a long enough evolutionary time or a large enough population size
\cite{Georgiou,Wittrup2000,Pluckthun}.

The Eigen \cite{ei71} and Crow-Kimura \cite{ck70}, or parallel, models
of viral quasispecies
evolution are among the simplest that capture the
basic processes of mutation, selection, and replication  that occur
in natural evolution.  These mathematical models exhibit
phase transitions, such that for mutation rates below critical
values, an identifiable quasispecies forms.  
The Eigen and parallel quasispecies models 
are archetypes of biological evolution, and they
have become a popular entry point to evolutionary biology for physicists 
\cite{Krug,Leuthausser,Tarazona,Peliti2002,bb97,sh04a,Deem2006b,Deem2006}.
Quantification of the mutational load of transfer  of genetic information
has been done by numerical solutions of
the single-mutation-per-replication Eigen model for the special case of
a linear replication rate function \cite{Levine}.  
It was found
that for intermediate population sizes and finite times, 
genetic transfer
dramatically speeds up the rate of evolution.
Phase diagrams were not determined, due to the focus on
finite times and population sizes.
We here derive by analytical calculation the mutational
load and evolutionary advantage induced by 
transfer of genetic information for arbitrary replication rate functions in 
both the parallel and continuous-time Eigen models
of quasispecies theory.  That is, we find the infinite-time,
infinite-genome-length,
and infinite-population-size phase diagrams and mean fitness
values of these models
of quasispecies evolution in the general case.
As an example, for the sharp-peak replication rate function,
transfer of genetic information has two effects: sharpening of the
population in the selected phase ($u=1$ instead of $0 < u \le 1$)
and maintaining the unselected phase as metastable to higher growth rates.

We model transfer of genetic information by
replacement of part of the genome in a random individual
with sequence taken at the same genomic location from
a random parent.
One common way for this process to occur in viruses or
bacteria is by recombination.  
We assume that each of these
exchanges of genetic information changes only
one base of the sequence, such as might occur by homologous recombination
in a stable population with relatively low diversity.
We consider an infinite population of individuals with
fixed genome size $N$, and each site of the genome
may be in one of two states (e.g.\ purine or pyrimidine).
A given sequence $i$ reproduces at rate
$r_i$, and point mutations occur at rate $\mu$ per site to change 
sequence $j$ to $i$.

Transfer of genetic information occurs in
the process in which a base at position
$k$ from any sequence $j$ 
randomly replaces the base at position $k$ in sequence $i'$
with frequency $\nu$.
Quasispecies theory with transfer of genetic information
is described by the equation
\begin{eqnarray}
 \frac{d q_i }{dt} &=& 
 r_i q_i
+ \sum_{j=1}^{2^N} \mu_{i j} q_j
+ \nu \frac{
\sum_{k=1}^N \sum_{i'}' \sum_j' q_{i'} q_{ j}
}{
 {\sum}_{m=1}^{2^N} q_m
}
- \nu N q_i
\nonumber \\ &=&
 r_i q_i
+ \sum_{j=1}^{2^N} \mu_{i j} q_j
+ \nu 
\sum_{k=1}^N
(q_i + q_{\sigma_1(k) i} )
\frac{
  \sum_j' q_j
}{
 {\sum}_{m=1}^{2^N} q_m
}
\nonumber \\ && 
- \nu N q_i \ ,
\label{310}
\end{eqnarray}
where the primes indicate that
sequence $j$ equals sequence $i$ at position $k$, and 
sequence $i'$ equals sequence $i$ except possibly
at position $k$.  
The notation $i'=\sigma_1(k) i$ indicates
the sequence $i'$ that results from changing base $k$ in sequence $i$.
We have defined $q_i$ to be
proportional to the probability of sequence $i$ in the population.
We note that the recombination term conserves particle
number: taking the sum of Eq.\ (\ref{310}) over $i$ causes the
mutation and recombination terms to cancel.
We define the average spin at position $k$ at time $t$ as
$u(k)$,
$u(k) = \sum_j (\delta_{s^j_k, +1} - \delta_{s^j_k, -1}) q_j / 
\sum_m q_m$, where $s^j_k$ represents the base at position $k$
of sequence $j$.
The recombination process is, thus, described by
\begin{eqnarray}
 \frac{d q_i }{dt} &=& 
 r_i q_i
+\mu \sum_{k=1}^{N}  \left( q_{\sigma_1(k) i} - q_i \right)
+ \nu 
\sum_{k=1}^N
(q_i + q_{\sigma_1(k) i} )
\nonumber \\
&& \times \left(
\frac{1+u(k)}{2} \delta_{s^i_k, +1} +
\frac{1-u(k)}{2}\delta_{s^i_k, -1}
\right)
- \nu N q_i  \ .
\nonumber \\ 
\label{321}
\end{eqnarray}

We analyze this equation at long times, when $u$ becomes
independent of both time and position.  In this limit,
the non-linear Eq.\ (\ref{321}) becomes a linear equation,
with a self-consistency condition for $u$.  We find the long-time
solution by a mapping
to a two-component field theory, using the procedure
of \cite{Deem2006b}.
We define the space
$ \vert \psi(t) \rangle = \sum_{i}^{2^N}
q_{i}(t)
\vert  s^i \rangle
$.
The evolution equation is then cast as
$
\frac{d}{dt} \vert \psi \rangle
= -\hat H \vert \psi \rangle
$.
The space is
represented in terms of creation and annihilation operators.
We seek the operator form of Eq.\ (\ref{310}).
  Each spin state, $ \vert s_n^i \rangle =
\vert +1 \rangle$,
$\vert -1 \rangle$,
 is created by a different operator.

We introduce two pairs of
creation and annihilation operators for the 
space, with the constraint
that at each position one and only one particle is present.
We label the creation operators at position $j$
as $ \hat {\vec a}^\dagger(j) =
(\hat a^\dagger_1(j),
\hat a^\dagger_2(j))$.
The Hamiltonian is given by
\begin{eqnarray}
-\hat H & = &
\mu 
\sum_{j=1}^N
 [\hat M(j) -1] 
 + \nu N \hat R
\nonumber \\ && +
N  
f\left[ \frac{1}{N} \sum_{j=1}^N \hat {\vec a}^\dagger(j)
\cdot  \sigma_3
\hat {\vec a}(j) \right] \ ,
\label{6}
\end{eqnarray}
where $\mu$ is the mutation rate,
$\hat M(n)$ mutates $s^i$ at position $n$,
$N f(u_i)=r_i$ is the replication rate,
$\nu N $ is the recombination rate,
and
$\hat R$ is the recombination operator.
The mutation operator $M(j)$ operates on site $j$ and
are defined by
$
\hat M(j) = 
{\hat {\vec a}}^\dagger(j) \cdot
\sigma_1
{\hat {\vec a}}(j)
$.
The recombination operator is
given by
\begin{eqnarray}
\hat R  &=& \frac{1}{N} \sum_{j=1}^N
[
a_1^\dagger(j) [a_1(j) + a_2(j)]
\frac{1+u}{2}
\nonumber \\ &&
+   a_2^\dagger(j) [a_1(j) +  a_2(j)]
\frac{1-u}{2}
  - 1
] \ .
\label{301a}
\end{eqnarray}

The formal solution is
$
\vert \psi(t) \rangle
= e^{ -\hat H  t} \vert \psi(0) \rangle
$,
which implies that the joint probability distribution at time $t$ is
given by
$
q_i (t) =
 \langle s_i \vert 
 e^{ -\hat H  t} 
\sum_{l}
q_{l}(0)
\vert  s^l  \rangle
$.
We introduce the coherent state representation
$
\vert \vec z \rangle = e^{\hat {\vec a}^\dagger \cdot \vec z -
{\vec z}^* \cdot \hat {\vec a} }
\vert 0,0 \rangle
$,
where $\vec z = (z_1, z_2)$ is a two-vector.
The coherent states satisfy a completeness relation
$
I = 
\int [{\cal D} {\vec z}^* {\cal D} {\vec z}]
\vert {\vec z} \rangle \langle {\vec z} \vert
$.
The overlap of the coherent states is given by
$
\langle {\vec z}' \vert {\vec z} \rangle =
e^{-\frac{1}{2} 
[ 
{\vec z}^{'*} \cdot ({\vec z}' - {\vec z}) -
( {\vec z}^{'*} - {\vec z}^* ) \cdot {\vec z}
]
}
$.
Using \cite{Deem2006b}, we 
use the Trotter factorization to find the
evolution operator is
\begin{eqnarray}
e^{-\hat H t} &=&
\lim_{M \to \infty}
\int [\prod_{k=0}^M {\cal D} {\vec z}^*_k {\cal D} {\vec z}_k]
\vert {\vec z}_M \rangle
\left[
\prod_{k=1}^M 
\langle {\vec z}_k \vert e^{- \varepsilon \hat H} \vert
{\vec z}_{k-1} \rangle 
\right]
\langle {\vec z}_0 \vert \ ,
\nonumber \\
\label{14}
\end{eqnarray}
where $\varepsilon = t/M$.

The
probability to go from an initial state $q_{(i)=\gamma}$ to a final
state $q_{(i' ) = \gamma'}$,
where $ 1 \le \gamma_j \le 2$ indicates the composition of the
pair of bases at position $j$, is
\begin{eqnarray}
 P &=& \lim_{M \to \infty}
\int {\cal D} \bar { \xi}^* {\cal D} {\xi}
e^{\varepsilon N \sum_{k=1}^M 
[ f(\xi_k) -  \bar \xi_k \xi_k ]
}
\prod_{j=1}^N
Q_{\gamma'_j \gamma_j}(j) \ ,
\nonumber \\
\label{28}
\end{eqnarray}
where
$
Q(j) = \prod_{k=1}^M [ I + \varepsilon B_k(j)]
$, with
$
B_k(j)  =
\mu (\sigma_1-I)
+
\nu (D - I)
+\bar \xi_c \sigma_3
$, and
$D = 
\left(
\begin{array}{cc} 
\frac{1+u}{2} & \frac{1+u}{2} \\
\frac{1-u}{2} & \frac{1-u}{2}
\end{array}
\right)
$.

We are interested in the probability distribution at long times, which for a
given $u$ grows as $e^{f_{\rm m} t}$
by the Perron-Frobenius theorem,
where $f_{\rm m}$ is the
largest eigenvalue of $- \hat H$, and $f_{\rm m}$
is equal to the mean replication
rate at long times \cite{Deem2006b}, when $u$ is self-consistently determined.
We evaluate the contribution to
this eigenvalue from $Q(j)$ by considering the expression
$
{\rm Tr}~ Q(j)
$.
We find 
\begin{eqnarray}
\ln {\rm Tr~} Q(j) \sim t \left(
\sqrt{(\mu + \nu/2)^2 +  \nu u \bar \xi_c + \bar \xi_c^2}
 - \mu - \nu/2  \right) \ .
\label{303}
\end{eqnarray}
We note that in the limit of infinite $\nu$, $\ln {\rm Tr}~ Q(j)
 \sim t  u \bar \xi_c$.
The mean replication rate is given by
$
f_{\rm m} = \max_{\xi_c, \bar \xi_c}
\{f(\xi_c)  - \bar \xi_c \xi_c + [\ln {\rm Tr}~ Q(j)]/t \}$.
Maximizing over $\bar \xi_c$, we find
\begin{eqnarray}
f_{\rm m} &=& \max_{\xi_c}
\bigg\{f(\xi_c)
+ \left[
(\mu+\nu/2)^2 - (\nu u /2)^2
\right]^{1/2}
\sqrt{1 - \xi_c^2}
\nonumber \\ &&
+ \nu u \xi_c/2 - \mu - \nu/2
\bigg\} \ .
\label{305a}
\end{eqnarray}
The observable surface magnetization, $u$,  is given by the implicit
self-consistency condition
$f(u) = f_{\rm m}$.
Thus, the two variables $\xi_c$ and $u$ need
to be determined when solving Eq.\ (\ref{305a}).  This procedure
provides the exact solution to the parallel model 
of recombination for a general replication rate function.

\begin{figure}[t!]
\epsfig{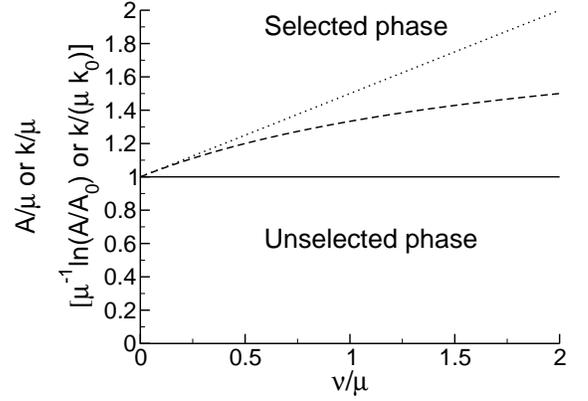}
\caption{The selected phase, in which a finite fraction of the 
population has a non-zero magnetization, is shown for the parallel model
with recombination.  The rate of genetic exchange is $\nu$, 
and the rate of mutation is $\mu$.
The phase diagram is shown for
the replication rate $f(u) = A \delta_{u,1}$  (solid line, with short-dashed
line the metastable limit) and
the replication rate $f(u) = k u^2/2$ (dashed line).
Also shown is the phase diagram for the Eigen model
for $f(u) = (A-A_0) \delta_{u,1} + A_0$ (solid line,
with short-dashed line the metastable limit)
and  $f(u) = k u^2/2 + k_0$ (long-dashed line)
[ordinate coordinates in brackets].
\label{fig1}
}
\end{figure}

To illustrate how recombination affects the
error-threshold phase transition, we calculate the error threshold
for three different replication rate functions.
We first consider in detail  $f(1) = A$ and $f = 0$
otherwise.  Eq.\ (\ref{305a}) is maximized at
$\xi_c = 1$ or $\xi_c = 0$.  
The error threshold is given for $u=0$ by $A > \mu + \nu/2$.
The self-consistency condition $f_{\rm m} = f(u)$ can
only be satisfied by $u = 1 - O(1/N)$.
Thus, due to the non-linearity, $u = 1$ in the selected phase, in contrast
to the case without recombination, for which
$u = 1 - \mu/A$ \cite{Deem2006b}.
In the selected phase 
$f_{\rm m} = A - \mu$.
Thus, the true error threshold is $A > \mu$,
with $A > \mu + \nu/2$ the limit of metastability for initial
conditions with $u \approx 0$.
If we
are in the selected phase and reduce the replication rate of the sharp peak,
we transform to the unselected phase at the solid line of Figure \ref{fig1}.
If, however, we are in the unselected phase and increase the replication rate
of the sharp peak, we may not transform to the selected phase until
reaching the short-dashed line of Figure \ref{fig1}.
We next consider in detail
the case of the quadratic replication rate $f(u) = k u^2/2$.
By setting Eq.\ (\ref{305a}) equal to $f(u)$ for small $u$,
we find that the error threshold is given by
$k > (\mu + \nu)/[ 1 + \nu/(2 \mu)]$.
At small $\nu$, recombination has again shifted the
transition by $+\nu/2$.
As an example of general parameter values, 
for $\nu = 1$, $\mu = 1$, and $k=2$, we find $u = 0.4671$ and
$f_{\rm m} = 0.2182$.
Solving Eq.\ (\ref{310}) numerically for
$N = 10^3$, $\nu = 1$, $\mu = 1$, and $k=2$, we find $u = 0.4662$ and
$ f_{\rm m} = \langle f(u_l) \rangle = 0.2183$.
Note recombination introduces a genetic load for the quadratic replication
rate, since in the absence of recombination $u = 1 - \mu/k = 1/2$ for
these parameters.
Note also that metastability does not occur for the quadratic replication
rate.
The error-threshold phase diagram for these two cases
is shown in Figure \ref{fig1}.
We finally consider in detail the linear fitness
$f(\xi) = k_0 + k \xi$.  We find that for all values of $k_0$, 
$k>0$, $\mu \ge 0$, and $\nu \ge 0$, the optimal value of $\xi_c$ is
positive.  Thus, the selected phase always occurs

We now turn to consider recombination in the Eigen model.
In the Eigen model, when a virus
reproduces, the virus copies its genome, making mutations
at a rate of $1-y$ per base per replication.  The
un-normalized probability
distribution in genome space satisfies
\begin{eqnarray}
\frac{d q_i}{dt}=
\sum_{j,k=1}^{2^N}\left[B_{ij} C_{jk} r_k-
\delta_{ij} \delta_{ik} D_i \right] q_k \ .
\label{701}
\end{eqnarray}
The degradation rate is defined analogously to the replication rate
by $D_i = N d(u_i)$.
Here the  transition rates are given by
$B_{ij}=y^{N-d(i,j)}(1-y)^{d(i,j)}$.
We define the parameter $\mu = N(1-y)/y$ to characterize
the per genome replication rate, where we take $\mu = O(1)$,
consistent with observed mutation rates in many
viruses and bacteria \cite{Drake1991}.
We also have
$
C_{jk} 
 \sim
 \bigg[
-\nu  + (\nu/N) \sum_{l=1}^N 
(\delta_{jk} + \delta_{\sigma_1(l)j, k})
\left(
\frac{1+u(l)}{2} \delta_{s^j_l, +1} +
\frac{1-u(l)}{2}\delta_{s^j_l, -1}
\right)
\bigg] 
$.
The rate of
genetic exchange per site is $\nu/N$.
This equation is generated to $O(N^0)$ by the Hamiltonian
\begin{eqnarray}
-\hat H & = &
N  
e^{-\mu + \frac{\mu}{N} \sum_{j=1}^N \hat {\vec a}^\dagger(j) \cdot
      \sigma_1 \hat {\vec a}(j)}
\times
e^{-\nu + \frac{\nu}{N} \sum_{j=1}^N \hat {\vec a}^\dagger(j) \cdot
      D \hat {\vec a}(j)}
\nonumber \\ &&
\times f\left[ \frac{1}{N} 
\sum_{j=1}^N 
\hat {\vec a}^\dagger(j) \cdot
\sigma_3 
\hat {\vec a}(j)
\right]
\nonumber \\ &&
- N d\left[ \frac{1}{N} 
\sum_{j=1}^N 
\hat {\vec a}^\dagger(j) \cdot
\sigma_3 
\hat {\vec a}(j)
\right] \ .
\label{702}
\end{eqnarray}
We define $\chi_k = \frac{1}{N} \sum_{j=1}^N
 \hat {\vec a}^\dagger(j) \cdot
      D \hat {\vec a}(j)$ and
$\eta_k = \frac{1} {N} \sum_{j=1}^N {\vec z}^*_{k} (j)
\sigma_1  {\vec z}_{k-1}(j)$.  We
integrate out the $\bf z$ field, to find
$
B_k(j) = \bar \eta_k \sigma_1 + \bar \xi_k \sigma_3 
+ \bar \chi_k D
$.
The action is, therefore, given to $O(N^0)$ by
\begin{eqnarray}
-S &=& \varepsilon N \sum_{k=1}^M
\bigg[
e^{-\mu + \mu \eta_k - \nu + \nu \chi_k} f(\xi_k) - d(\xi_k)
\nonumber \\ &&
-\bar \xi_k \xi_k 
-\bar \eta_k \eta_k
-\bar \chi_k \chi_k
\bigg] + N \ln {\rm Tr}~ Q(j) \ ,
\label{704}
\end{eqnarray}
where $\ln {\rm Tr}~ Q(j) \sim t ( \sqrt{(\bar \eta_c + \bar \chi_c/2)^2 +
u \bar \chi \bar \xi_c + \xi_c^2}) + \bar \chi_c/2$.
Note that the probability of any of the $N$ bases undergoing both
mutation and recombination is $O(\nu \mu /N)$.
We have $
f_{\rm m} = \max_{\xi_c, \bar \xi_c, \eta_c, \bar \eta_c,
\chi_c, \bar \chi_c}
\{
e^{-\mu + \mu \eta_c - \nu + \nu \chi_c
} f(\xi_c) - d(\xi_c)
  - \bar \xi_c \xi_c -\bar \eta_c \eta_c
-\bar \chi_c \chi_c
+ [\ln {\rm Tr}~ Q(j)]/ t \}
$.
Maximizing over $\bar \xi_c$, $\bar \eta_c$, and
$\bar \chi_c$, we find
that $\bar \eta_c \eta_c + \bar \chi_c \chi_c + \bar \xi_c \xi_c
= (\ln q)/t$.
We use the additional relation $\mu \bar \chi_c = \nu \bar \eta_c$ to
find
$\eta_c(\xi_c) = \{
(1-\xi_c^2)/
[1 - \nu^2 u^2 / (2 \mu + \nu)^2 ] \}^{1/2}$ and
$\chi_c(\xi_c) =
\left[
1 + \eta_c + u \xi_c - 
u \left(
\eta_c^2 + \xi_c^2 - 1
\right)^{1/2}
\right]/2$.
The mean replication rate is given by the expression
\begin{eqnarray}
f_{\rm m} &=& \max_{\xi_c}
\left\{
e^{-\mu + \mu \eta_c - \nu + \nu \chi_c
} f(\xi_c) - d(\xi_c)
\right\} \ .
\label{707}
\end{eqnarray}
The observable $u$ is given implicitly by
$f_{\rm m} = f(u) - d(u)$.

To illustrate how recombination during
replication affects the
error-threshold phase transition, we calculate the error threshold
for three different replication rate functions.
We first consider in detail  $f(1) = A$ and $f = A_0$
otherwise.  Eq.\ (\ref{707}) is maximized at
$\xi_c = 1$ or $\xi_c = 0$.
The error threshold is given for $u=0$ by the equation
$A e^{-\mu - \nu/2} > A_0$.  
Due to the non-linearity, $u = 1$ in the selected phase, in contrast
to the case without recombination where
$u = (A e^{- \mu} - A_0)/(A-A_0)$ \cite{Deem2006b}.
The mean replication rate is given by
$f_{\rm m} = A e^{-\mu}$.
Thus,  the true error threshold is $A e^{-\mu} > A_0$,
with $A e^{-\mu - \nu/2} > A_0$ the limit of metastability for initial
conditions with $u \approx 0$.
The limits of the bistable region, which is reminiscent of the bistable
region found numerically in
a related Eigen model with a type of recombination \cite{Nowak},
are demarked by the solid and short-dashed lines in 
in Figure \ref{fig1}.
For our second example, we consider the
quadratic fitness $f(\xi) = k_0 + k \xi^2 / 2$.
By setting Eq.\ (\ref{707}) equal to $f(u)$ for small $u$,
we find that the error threshold is given by
$k > k_0 (\mu + \nu)/[ 1 + \nu/(2 \mu)]$.
At small $\nu$, recombination has again shifted by 
transition by $+\nu/2$.
The error-threshold phase diagram for the sharp peak and
quadratic replication rate cases
is shown in Figure \ref{fig1}.
For our third example, we consider in detail the linear fitness
$f(\xi) = k_0 + k \xi$.  We find that for all values of
$k>0$, $\mu \ge 0$, and $\nu \ge 0$, the optimal value of $\xi_c$ is
positive, and the selected phase always occurs.

How is it that recombination changes the phase diagram or mean fitness?
After all, this process simply replaces an allele with another allele
randomly chosen from the distribution of alleles at that site.
This process, however, reduces the correlations between the
composition of alleles at different sites in the sequence.  These correlations
are non-zero \cite{Deem2006b}, and so their
reduction changes the dynamics and the steady-state distribution.
Recombination sharpens
the phase transition for $f(\xi) = f(0) + \delta_{\xi,1} \Delta A$,
turning it into
step function for $u$.
More generally, recombination propagates favorable
mutations throughout the population, thereby typically increasing the
rate of evolution.

We may alternatively interpret each site in our model as
a coarse-grained representation of an allele or gene, each
with only two states that may be changed either
by mutation or gene transfer.
We may also consider that the genome
has been ordered  so that all the mobile genetic elements
are collected, say, at the end.  If the genome can be considered
approximately constant in length and if the replication rate can 
be approximately expressed by the quasispecies
assumption $r_i = N f(u_i)$, then the model we have discussed is
a representation of horizontal gene transfer in a population
of evolving bacterial species, because homology is not a 
prerequisite for horizontal gene transfer in nature or our model.
The diversity in the population
represents the species diversity in a bacterial order, 
family, or genus.
As with natural gene transfer by mobile elements, our
model does not assume homology is required.
Our results make a couple generic predictions:
1) for a sharp peak fitness, such as might be induced by
a novel antibiotic, a population with horizontal gene transfer 
tends to be more uniformly resistant ($u \approx 1$) than is
a population without ($ 0 < u < 1$),
2) while the contributions of mutation and horizontal gene transfer
to the mean fitness are not identical, in Eq.\ (\ref{305a}) or (\ref{707}),
they are similar for smooth replication rates.
Horizontal gene transfer tends to incorporate alleles with new function,
whereas mutation tends to adapt existing alleles for improved
function.  The observed rates of
horizontal gene transfer and mutation might be expected
to be the same order of magnitude, therefore, to balance the resources
expended on the complementary tasks of
large-scale evolution and local adaptation.
As an example, we consider the evolution of \emph{E.\ coli} from
\emph{Salmonella}.  The rate of evolution due to horizontal gene transfer
is estimated to be 16,000 bases/million years,
while that due to point mutation, is 22,000 bases/million years
\cite{Lawrence1998}.  These are observed rates, and so
selection plays a role.  The underlying rate of horizontal
gene transfer in \emph{E.\ coli} has been estimated to be
about $10^{-6}$ genes per cell per replication \cite{Berg2003},
which corresponds to a change of roughly $10^{-3}$ bases per
sequence per replication, given the average \emph{E.\ coli} gene length
of $10^3$ bases.  Taking the typical underlying \emph{E.\ coli} 
mutation rate of $5 \times 10^{-10}$ per base per replication \cite{Drake1991}
and noting that the \emph{E.\ coli} genome length is
$\approx 5 \times 10^6$ bases, we find 
that point mutation modifies approximately
$2.5 \times 10^{-3}$ bases per sequence per replication.
Interestingly, this same equality of underlying 
horizontal gene transfer and
point mutation rates per base per replication is also observed in quantitative
models of laboratory directed protein evolution optimized for evolutionary
rate \cite{Deem1999}.

\bibliography{hgt}

\end{document}